\documentclass[12pt,preprint]{aastex}
\newcommand{\beq}{\begin{equation}}
\newcommand{\beqa}{\begin{eqnarray}}
\newcommand{\eeq}{\end{equation}}
\newcommand{\eeqa}{\end{eqnarray}}
   
\newcommand{\etal}{{\it et al. }}

\newcommand{\lsim}{\la}
\newcommand{\gsim}{\ga}

\newcommand{\lmk}{\left(}
\newcommand{\rmk}{\right)}
\newcommand{\lnk}{\left\{ }

\newcommand{\rnk}{\right\} }

\newcommand{\lla}{\left\langle}
\newcommand{\p}{\partial}
\newcommand{\rra}{\right\rangle}
\shorttitle{}
\shortauthors{Takahashi \& Seto}

\begin{document}

\title{
Parameter estimation for  Galactic binaries
 by LISA
}
\author{Ryuichi Takahashi}
\affil{
Department of Physics, Kyoto University,
Kyoto 606-8502, Japan 
}
\author{Naoki Seto}
\affil{
Department of Earth and Space Science, Osaka University,
Toyonaka 560-0043, Japan
}


\begin{abstract}
We calculate how accurately parameters of the short-period binaries
 $(10^{-4}~{\mbox Hz} \lsim f\lsim 10^{-2}~{\mbox Hz})$
 will be determined from the gravitational waves   by LISA.
In our analysis   the chirp signal ${\dot f}$ is newly included as a
 fitting parameters and dependence on observational period  or wave
 frequency is studied in detail. Implications for gravitational
wave astronomy are also discussed quantitatively.
\end{abstract}



\section{Introduction}

The Laser Interferometer space Antenna (LISA), a joint project of 
NASA
and ESA (European Space Agency) would establish
gravitational wave astronomy at low frequency band $(10^{-4}~{\mbox 
Hz}
\lsim f\lsim 10^{-1}~{\mbox Hz})$.   It would bring us essentially 
new
information of the  Universe (Bender et al. 1998). For example
gravitational waves from 
merging super massive black holes (SMBHs) would be detected with
significant 
signal-to-noise ratio ($SNR$) $\gsim 10^3$,  though event rate of 
 such merging is highly unknown  ({\it e.g.}
Haehnelt 1994,  Vecchio 1997). Galactic binaries are promising sources 
of
LISA (Mironowskii 1965, 
Evans, Iben \& Smarr 1987, Hils, Bender \& Webbink 1991, Webbink \& Han
1998).  
Gravitational
waves from some known binaries ({\it e.g.} X-ray 
binary 4U1820-30) would  be detected with $SNR>5$ by one year
integration (Bender et al. 1998). In addition more than thousands of 
close
white dwarf binaries (CWDBs) are expected to exist in LISA band (for
recent studies see Yungelson et al 2001, Nelemans et al. 2001,
Napiwotzki et al. 2002). 
Our  target in this article is these Galactic binaries. We examine 
how
accurately information of binaries can be extracted from 
gravitational
waves observed  by LISA.

Cutler (1998) studied the estimation errors   for binary parameters 
with
special attention to  angular variables  such as direction and
orientation of binaries (see also Peterseim et al. 1997; Vecchio \&
Cutler 1998; Cutler \& Vecchio 1998;  Hughes 2002; Moore \&  
Hellings 2002).
He used approximation that emitted 
gravitational waves would be  monochromatic, namely
neglected the effects of the chirp signal $\dot f$.
But the wave frequency or the chirp signal are fundamental quantities
for gravitational wave astronomy. From the measured chirp signal
 $\dot f$ we can obtain the so called chirp mass
$M_c=M_1^{3/5}M_2^{3/5}(M_1+M_2)^{-1/5}$  ($M_1, M_2$: masses of two
stars) for a binary whose orbital evolution is determined by
gravitational radiation reaction. Furthermore the distance to the 
binary
could be estimated from the chirp signal $\dot f$ and the amplitude of
the wave signal (Schutz 1986). 
The frequency $f$ itself  also contains important information. One of 
the
authors  (Seto 2001) pointed out that signature of the periastron 
advance
could be detected in gravitational waves  from an eccentric binary by
measuring its wave frequencies preciously. If this method works well,
 we can estimate
the total 
mass $M_{total}=M_1+M_2$ of the binary beside the chirp mass, and
consequently 
each mass of the binary is obtained separately.

Estimation errors for fitting parameters
correlate  complicatedly to each other and
 depend largely on observational situations. For example longer
observational periods  would improve not only signal to noise ratio but
also
resolution 
of the frequency space. Note that the latter is crucial for reducing 
Galactic
binary confusion noise, as the number of resolved
binaries increases with decrease of the frequency bin $\propto T_{obs}^
{-1}$ 
(see Seto 2002 for details). 
At LISA age it would become an interesting observational
challenge 
to  
optically  identify the binaries whose gravitational waves are 
detected
by LISA. We also discuss impacts of these observational efforts on
 estimation of binary parameters.

The magnitude of
the chirp 
signal $\dot f$ has strong dependence on   the wave frequency $f$ as
${\dot f}\propto M_c^{5/3}f^{11/3}$.  At lower frequencies the
estimation error 
$\Delta {\dot f}$ would be larger than the signal $\dot f$ itself.  
Then
it would be better to remove  $\dot f$ from  fitting parameters 
and
simply 
put ${\dot f}=0$ from the beginning. By evaluating the estimation 
error
$\Delta {\dot f}$ we can quantitatively  discuss these prescriptions 
for
 signal  analysis.

In this article we only study parameter estimation errors on the
assumption that (i) signal has been detected with high SNR, and (ii)
noises are Gaussian distributed. We evaluate the estimation errors 
using
the Fisher 
information matrix for maximum likelihood method. For low SNR our 
simple
analysis would not be valid,  as the probability distribution function 
of the
fitting parameters could become highly complicated ({\it e.g.}
multimodal) (Balasubramanian \&
Dhurandhar 1998). Thus our result should
be regarded as a lower bound of the estimation errors. Non-Gaussian
nature of noises would farther increases 
errors. This is a very important problem, but detailed quantitative
analysis would   be a formidable task. 

This article is organized as follows. In \S 2 we  briefly discuss 
the gravitational waveforms of chirping binaries  and the data stream
obtained by LISA. Then we mention the parameter estimation  based
on the Matched filtering analysis.  In \S 3 we numerically evaluate 
the 
parameter estimation errors, and discuss its dependence on 
observational
period or wave frequency in detail. 
\S 4 is devoted to  summary and discussions.

\section{Gravitational Waveforms and Parameter Extraction}

\subsection{Gravitational Wave Measurement with LISA}
LISA consists of three  spacecrafts forming an equilateral
 triangle, and orbits around the Sun, trailing $20^{\circ}$ behind the
Earth. 
The sides of the triangle are $L=5 \times 10^6$ km in length, and the
plane of 
 triangle is inclined at $60^{\circ}$ with respect to the ecliptic.
The triangle rotates annually. 
The gravitational wave signal is reconstructed from the three data
streams  that effectively correspond to three
time-varying  
 armlength data, ($\delta L_1, \delta L_2, \delta L_3$) for 
gravitational
 waves with $\lambda \lsim L$.
We basically analyze two data streams given by 
 $s_{I}(t)=(\delta L_1(t)-\delta L_2(t))/L$ and $s_{II}(t)=
 (\delta L_1(t)+\delta L_2(t)-2\delta L_3(t))/\sqrt{3}L$.
These data can be  regarded as the response of two
 $90^{\circ}$-interferometers  rotated by $45^{\circ}$ to one
another (Cutler 1998).   
The data  $s_{I,II}(t)$ contain both  gravitational waves signal
$h_{I,II}(t)$  to be fitted  by matched filtering and   the  noise
$n_{I,II}(t)$. 
The latter is constituted by the detectors noise
and the binary confusion noise.
As in  Cutler (1998) we assume that the noises are stationary, Gaussian
and uncorrelated with 
each  other.

The gravitational wave 
signals  $h_{I,II}(t)$ from a binary  are written as 
\beq
  h_{I,II}(t)=\frac{\sqrt{3}}{2} \left[ F^{+}_{I,II}(t)h_+(t)
 +F^{\times}_{I,II}(t)h_{\times}(t) \right],
\eeq
where $F^{+,\times}_{I,II}(t)$ are the pattern functions which depend 
on
the 
 source's angular position of the binary ($\bar{\theta}_S,\bar{\phi}_S$)
, its
 orientation ($\bar{\theta}_L,\bar{\phi}_L$) and detector's 
configuration. 
The angular variables with bars  are defined  in a fixed  barycenter
frame of 
 the solar system, and the  direction and orientation of the 
binary  are assumed to
 be constant  during the observation in this frame. The quantities
$h_{+,\times}(t)$ are the two polarization modes of 
 gravitational radiation  from  the  binary.  
 We can estimate both
($\bar{\theta}_S,\bar{\phi}_S$) and ($\bar{\theta}_L,\bar{\phi}_L$) 
from
the time profiles  of the two  signals due to LISA's annual  rotation
and revolution. 
 Further discussion and details about the pattern functions are seen in
 Cutler (1998).  We basically use his formulation but newly
 include the effects of
 the chirp signal $\dot f$.

\subsection{Gravitational Waveforms}
We study   short-period
 ($10^{-4}~{\rm Hz}\lsim f \lsim 10^{-2}~{\rm Hz}$) 
binaries such as the close white dwarf 
binaries 
 (CWDBs) or the neutron star binaries (NBs) in our  Galaxy   
 (Hils, Bender \& Webbink 1991). 
We only discuss binaries with circular orbits. This is an excellent
approximation for CWDBs, as their orbits are circularized by strong
 tidal interaction in their earlier evolutionary stages.
   At LISA band some NBs or neutron star-white dwarf
binaries would 
have non-negligible eccentricities $e\sim 0.1$ induced at their
  supernova
 explosions (see {\it e.g.} Brown, Lee, Zwart, \& Bethe  2001).   But
extension 
to these eccentric binaries is straightforward.

 The chirping 
gravitational waveform  is given  by the quadrupole
 approximation \cite{peters64} as 
\beqa
  h_{+}(t) &=& A \cos \left[ 2 \pi \left( f + \frac{1}{2} {\dot f} t 
 \right) t + \phi_D(t) + \phi_0 \right] \times \left[ 1+ \left( {\hat L
}
 \cdot {\hat n} \right)^2 \right],   \nonumber \\ 
  h_{\times}(t) &=& -2 A \sin \left[ 2 \pi \left( f + \frac{1}{2} {\dot
 f}
 t \right) t + \phi_D(t) + \phi_0 \right] \times \left( {\hat L} \cdot
 {\hat n} \right),
\label{wf}
\eeqa   
where ${\hat L}$ (given by $\bar{\theta}_L,\bar{\phi}_L$) is the unit
vector in the direction of 
 the binary's orbital angular
 momentum, ${\hat n}$ (given by $\bar{\theta}_S,\bar{\phi}_S$) is the
unit  vector toward the binary 
 and $\phi_0$ is an integral  constant. 
We regard the frequency $f$ and its time variation
${\dot f}$ as constants in the above equations.
A purely  monochromatic waveform has ${\dot f}=0$. This is the case 
 studied by 
 Cutler (1998).
When gravitational radiation reaction
 dominates evolution of the binary as in the
case of CWDBs or NBs, the  chirp signal ${\dot f}$ is given as
 ${\dot f}=( 96 \pi^{8/3} /5 )
f^{11/3} M_c^{5/3}$ with  the chirp mass $M_c$.
The perturbative expansion for the intrinsic 
evolution of the wave frequency 
 in equation (\ref{wf}) are valid under the condition 
 ${\dot f} T_{obs} \ll f$, where $T_{obs}$ is the observational period.
This condition is expressed as
\beq
 f \ll 0.14  \left( \frac{T_{obs}}{10 \mbox{yr}} 
 \right)^{-3/8} \left( \frac{M_c}{1 M_{\odot}} \right)^{-5/8}~\mbox{Hz}.
\eeq  
The amplitude $A$ in equation (\ref{wf}) is given in terms of the wave
frequency  $f$, chirp signal  $\dot f$ and the distance $D$ as
\beq
 A=\frac{5}{96 \pi^2} \frac{\dot f}{f^3 D}.
\label{amp}
\eeq
Thus we could determine the distance $D$, if we could measure three
observables $f$, $\dot f$ and $A$ (Schutz 1986). This is an important
aspect of gravitational wave astronomy.

The term  $\phi_D(t)$ in equation (\ref{wf}) is caused by  revolution 
of
LISA around the Sun and called the Doppler phase.
Its explicit form  is given by
\beq
  \phi_D(t)=2 \pi f R \sin \bar{\theta}_S \cos \left[ \bar{\phi}(t)-
 \bar{\phi}_S \right],
\label{dop}
\eeq
where $R=1$ AU and $\bar{\phi}(t)=2\pi t/T~(T=1 \mbox{yr})$ is the
direction of LISA in the fixed barycenter frame.

\subsection{Parameter Extraction}
Let us briefly discuss the matched filtering analysis and the parameter
estimation errors \cite{finn92,cf94}. We assume that
 the signal $h_\alpha(t)$ is characterized by some
unknown parameters $\gamma_i$ (eight  parameters in the present case:
$(A,f,{\dot f},\phi_0,{\bar \theta_S},
 {\bar \phi_S},{\bar \theta_L},{\bar \phi_L})$).
In the matched filtering analysis the variance-covariance matrix of the
parameter estimation error $\Delta 
\gamma_i$ is given by   inverse of the Fisher 
information matrix $\Gamma_{ij}$ as 
$\langle \Delta  
 \gamma_i \Delta \gamma_j \rangle = \left( \Gamma^{-1} \right)_{ij}$.
For a quasi-monochromatic binary (${\dot f}T_{obs}\ll f$) the noise
spectrum $S_n(f)$ is nearly constant in the frequency region swept by
the binary and the Fisher matrix simply becomes \cite{cutler98}
\beq 
  \Gamma _{ij} 
 = \frac{2}{S_n(f)} \sum_{\alpha=I,II} \int_0^{T_{obs}} dt~
 \frac{\partial h_{\alpha}(t)}{\partial \gamma_i}
 \frac{\partial h_{\alpha}(t)}{\partial \gamma_j}.
\label{fis}
\eeq

The  error boxes for the angular parameters ${\hat L}$ and ${\hat n}$
become
ellipses  in the celestial sphere  due to the correlation of 
two parameters $\theta$ and $\phi$. 
In this article we represent the estimation errors for direction and
orientation of binaries in the form defined in Cutler (1998) as
follows 
\beq
 \Delta \Omega = 2 \pi \sqrt{  \langle \Delta \mu^2\rangle
\langle \Delta \phi^2\rangle   
  -   \langle \Delta \mu \Delta \phi \rangle^2 },
\eeq    
where we have defined $\mu=\cos \theta$.
In the same manner the signal to noise ratio ($SNR$) is given by
\beq
  (SNR)^{~2} =  \frac{2}{S_n(f)}\sum_{\alpha=I,II}  \int_0^{T_{obs}} dt
~
h_{\alpha}(t)  h_{\alpha}(t). 
\label{snr}
\eeq

  From equations (\ref{fis}) and (\ref{snr}) it is apparent that the
expressions for the 
 estimation errors  $\lla \Delta \gamma_i\Delta \gamma_j\rra$
 do not depend on the noise spectrum $S_n(f)$ when they are normalized
by the signal-to-noise ratio \citep{cutler98}.  In this article we
extensively use this  normalization method. For example the parameter
estimation 
errors for a simple wave form $h(t)=A \sin[2\pi(f+{\dot f}t/2)t+\phi_0]
 $ 
with four  fitting parameters $(A,f,{\dot f},\phi_0)$ is easily
evaluated as in Seto (2002). The explicit forms for $\Delta A, \Delta f
$ and
$\Delta \dot f$ are given by 
\beqa
\frac{\Delta A}{A} &=& \frac{1}{SNR} =
0.1 \lmk \frac{SNR}{10} \rmk^{-1}, \label{aa} \\ 
\Delta f&=&\frac{4\sqrt3}{\pi}\frac{T^{-1}_{obs}}{SNR}=0.22 \lmk
\frac{SNR}{10}\rmk^{-1} T_{obs}^{-1},\label{af}\\
\Delta {\dot f}&=&\frac{6\sqrt5}{\pi}\frac{T^{-2}_{obs}}{SNR}=0.43 \lmk
\frac{SNR}{10}\rmk^{-1} T_{obs}^{-2}.\label{ac}
\eeqa
This simple analysis does not include information of the angular
parameters, but 
would be helpful to understand
 more detailed numerical analysis in the following
section. 
\section{Results}

\subsection{General Behavior}

We have numerically evaluated the uncertainties of the estimated
 parameters for  various
quasi-monochromatic binaries. In this subsection
 we show results for  a typical example with a fixed set of  
angular parameters  at $\cos \bar{\theta}_S=0.3,\bar{\phi}_S=5.0,
 \cos \bar{\theta}_L=-0.2 $ and $ \bar{\phi}_L=4.0$.
In Table \ref{t1} we show LISA's measurement accuracy for parameters 
  $(A,f,{\dot f},\Omega_S,\Omega_L)$  at  
frequencies $f=10^{-4}, 10^{-3}$ and $10^{-2}$Hz. We present our 
results
for two   observational
periods,  $T_{obs}=1$ and 10yr. 
All results are normalized by $SNR=10$ after integration over each
 observational period $T_{obs}$. 
These results simply scale as  $(SNR/10)^{-1}$ for errors $\Delta A, 
 \Delta f$ and $\Delta {\dot f}$ and $(SNR/10)^{-2}$ for error
 ellipses $\Delta \Omega_{S,L}$.  
The first row for each observational period 
 $T_{obs}$ represents  the case when all the eight
parameters $(A,f,{\dot f}, \phi_0,\bar{\theta}_S,\bar{\phi}_S,
 \bar{\theta}_L,\bar{\phi}_L)$ are 
included in the matched filtering analysis. The second row corresponds
to the case without the chirp signal $\dot f$.  Results
for $T_{obs}=1$yr are obtained under the same condition with
 Table 1 (case A) in Cutler (1998). Our numerical values  completely
 coincide with his ones.  The third row is the case 
 when direction  
 of the binary ($\bar{\theta}_S,\bar{\phi}_S$) 
is given exactly by other
method ({\it e.g.} optical identification of the binary) and 
removed from  the fitting parameters.

\begin{figure}
  \vspace{0.1cm}
  \hspace{3.75cm}
  \includegraphics[height=7.5cm,clip]{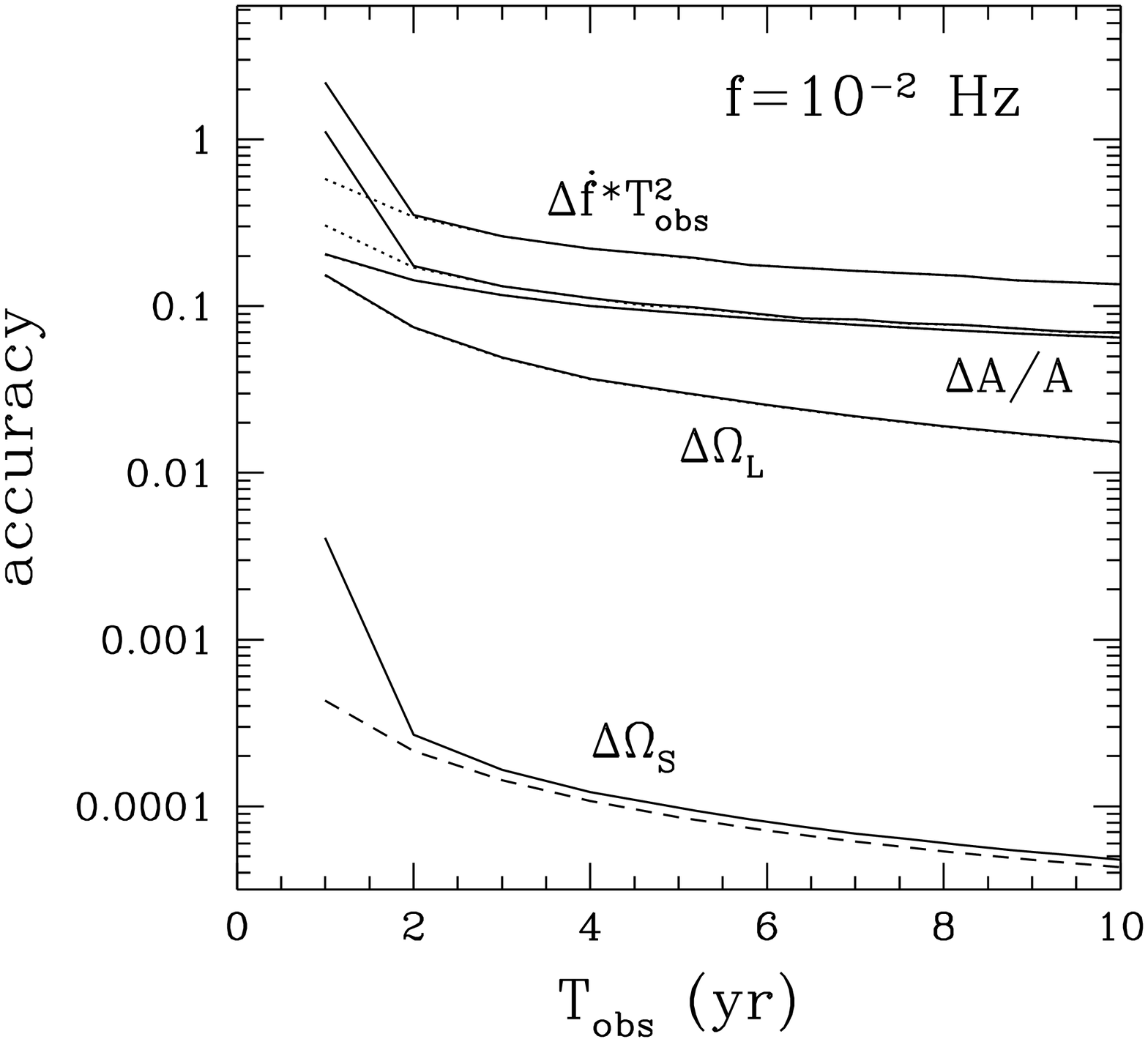}
  \vspace{0.1cm}
    \\
  \begin{minipage}[t]{7.5cm}
    \vspace{0.1cm}
     \includegraphics[height=7.5cm,clip]{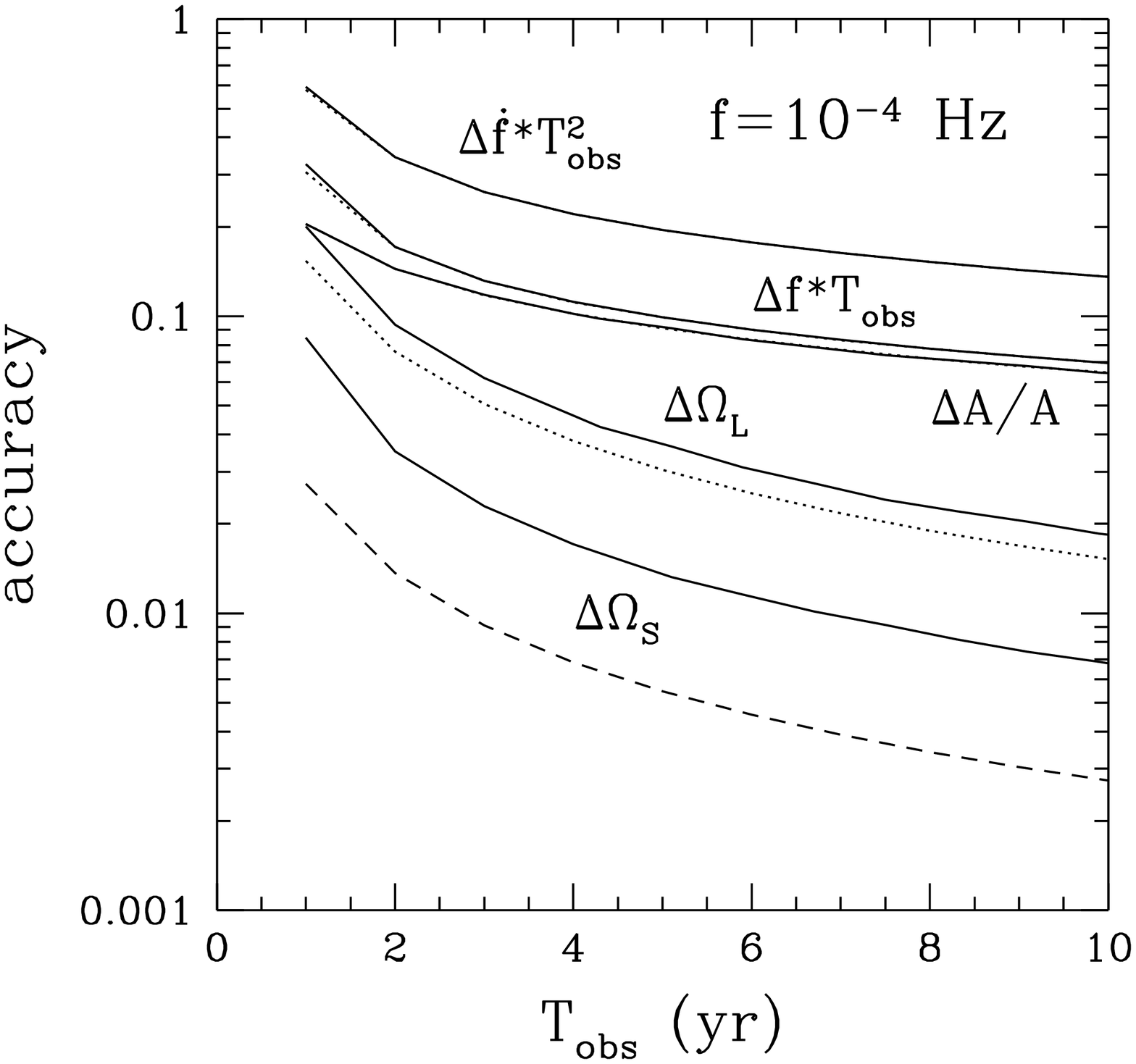} 
    \vspace{0.1cm}
  \end{minipage}
  \begin{minipage}[t]{7.5cm}
    \vspace{0.1cm}
     \includegraphics[height=7.5cm,clip]{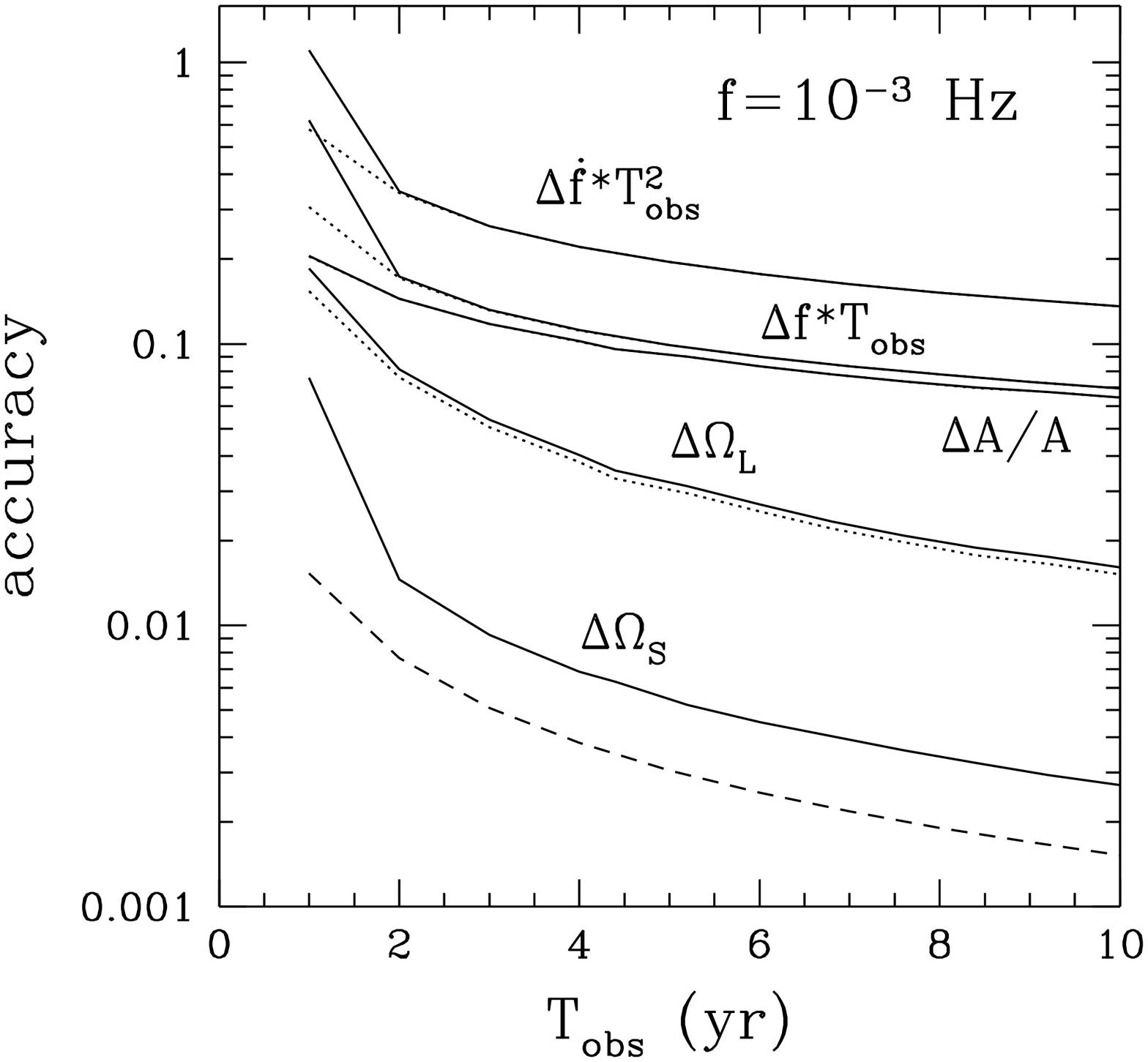}  
    \vspace{0.1cm}
  \end{minipage}
    \\
  \caption{LISA's measurement accuracy for the binaries
 as a function of the observational period $T_{obs}$ with angular
 parameters 
 $(\cos \bar{\theta}_S=0.3, \bar{\phi}_S=5.0, \cos \bar{\theta}_L=-0.2,
 \bar{\phi}_L=4.0)$. 
The solid lines correspond to $\Delta {\dot f},
 \Delta f, \Delta A,
 \Delta \Omega_L$ and $\Delta \Omega_S$ from top to bottom.
 The dotted lines represent the case source positions $(\bar{\theta}_S,
 \bar{\phi}_S)$ are removed from the fitting parameters,
 and the dashed lines represent the
 case the only source positions are included in the fitting parameters. 
The accuracies are normalized by $SNR=10$ at 1yr
 observation. 
} 
\label{f1}
\end{figure}

Figs 1 and 2 show clearly  dependence of the orbital parameters
 on wave frequency and observational period $T_{obs}$.
In Fig.\ref{f1} we show LISA's measurement accuracy as a function of
 observational period $T_{obs}$ at given frequencies
 $f=10^{-4},10^{-3}$ and $10^{-2}$ Hz.
All results are normalized by  $SNR=10$ at  integration period 
$T_{obs}=1$ yr. The solid lines are results for  fitting all the eight
parameters.
The dotted lines represent the case when the angular position
 $(\bar{\theta}_S,\bar{\phi}_S)$ are removed from the fitting
 parameters (see also Hughes 2002).
The dashed lines are results with fitting  only the angular position
 $(\bar{\theta}_S,\bar{\phi}_S)$.
For observational period  $T_{obs} \gsim 2$ yr, the
 difference between the solid and the dotted lines is very small
 especially for $\Delta A$, $\Delta f$ and $\Delta {\dot f}$
 irrespective of the frequency. Thus optical determination of the 
source direction $\Omega_S$ would only
slightly 
reduce the estimation errors of other parameters.
Asymptotic behaviors of errors $\Delta A$, $\Delta f$ and $\Delta {\dot
f}$ are given by
\beqa
\frac{\Delta A}{A} &=& 0.20 \lmk \frac{SNR}{10} \rmk^{-1} \label{ga},\\ 
\Delta f&=& 0.22 \lmk
\frac{SNR}{10}\rmk^{-1} T_{obs}^{-1} \label{gf}, \\
\Delta {\dot f}&=&0.43 \lmk
\frac{SNR}{10}\rmk^{-1} T_{obs}^{-2} \label{gc}, 
\eeqa 
(see also Table.1), and
 the asymptotic time dependence is given as
 $\Delta A \propto T_{obs}^{-1/2}, 
 \Delta f \propto T_{obs}^{-3/2},$ and $ \Delta {\dot f} \propto
 T_{obs}^{-5/2}$, since we have $SNR \propto T_{obs}^{1/2}$ from 
equation
 (\ref{snr}).
We have also
 $\Delta \Omega_{S,L} \propto SNR^{-2}\propto T_{obs}^{-1}$ for
directions and orientations of binaries. Numerical
 results for
$\Delta f$ and $\Delta {\dot f}$ in equations (\ref{gf}) and (\ref{gc})
 are almost identical to analytical ones in equations (\ref{af}) and
 (\ref{ac}). 
Hence we can expect that the information of the angular parameters  
 (the direction and orientation) does not affect the accuracy of
 the estimation for the frequency $f$ and the chirp signal ${\dot f}$
 for observational period $T_{obs} \gsim 2$ yr. 
But for the amplitude $A$, the above result (in [\ref{ga}]) is
 two times as large as  equation (\ref{aa}).
This is due to the fact that the amplitude $A$ is tied with the
 angular parameters (the inclination ${\hat L} \cdot {\hat n}$) in
 equation (\ref{wf}) and the estimation error $\Delta A$
 strongly depends on the error in ${\hat L} \cdot {\hat n}$
 (see the detailed discussion in the next subsection).

\begin{figure}
  \plottwo{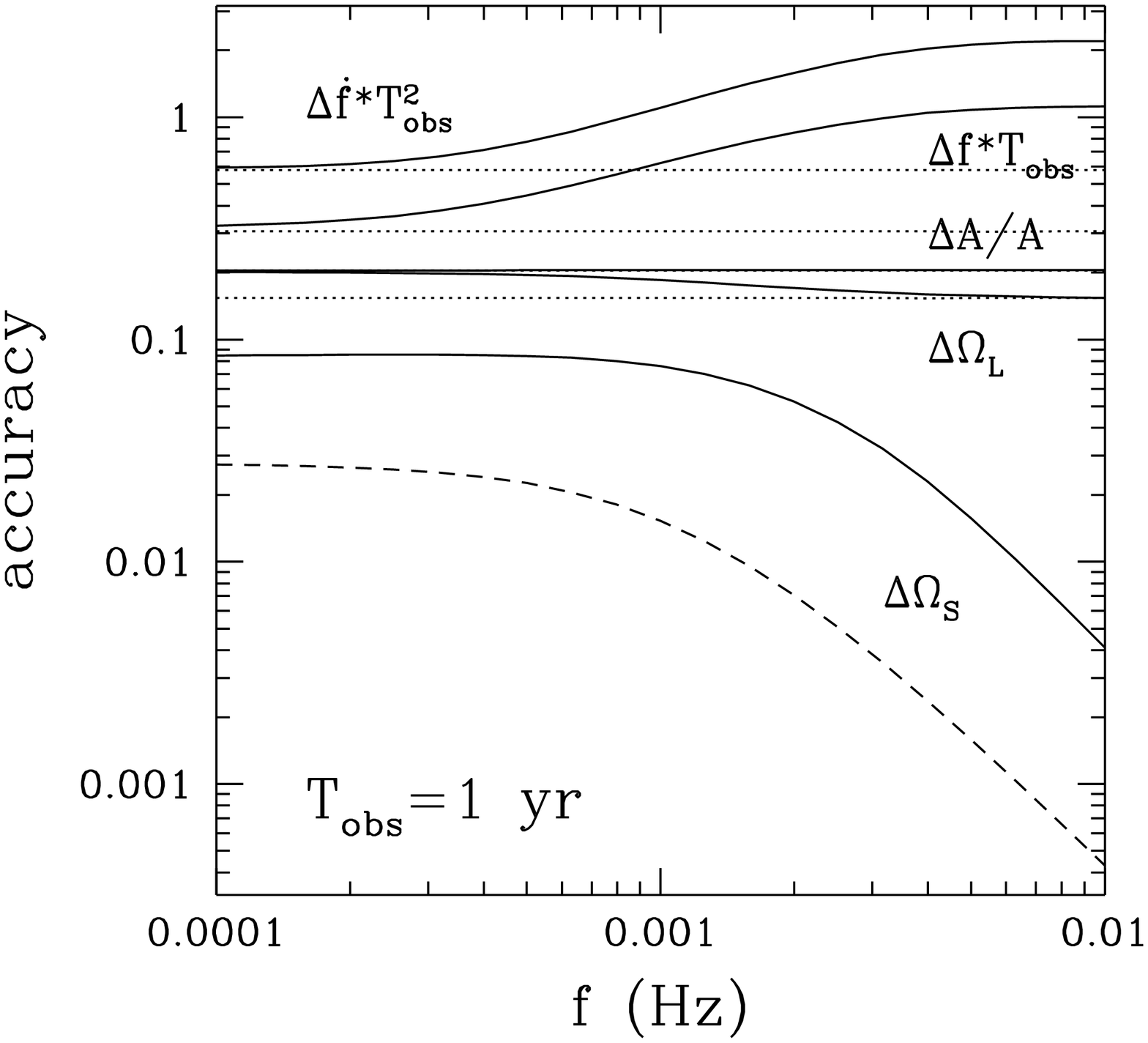}{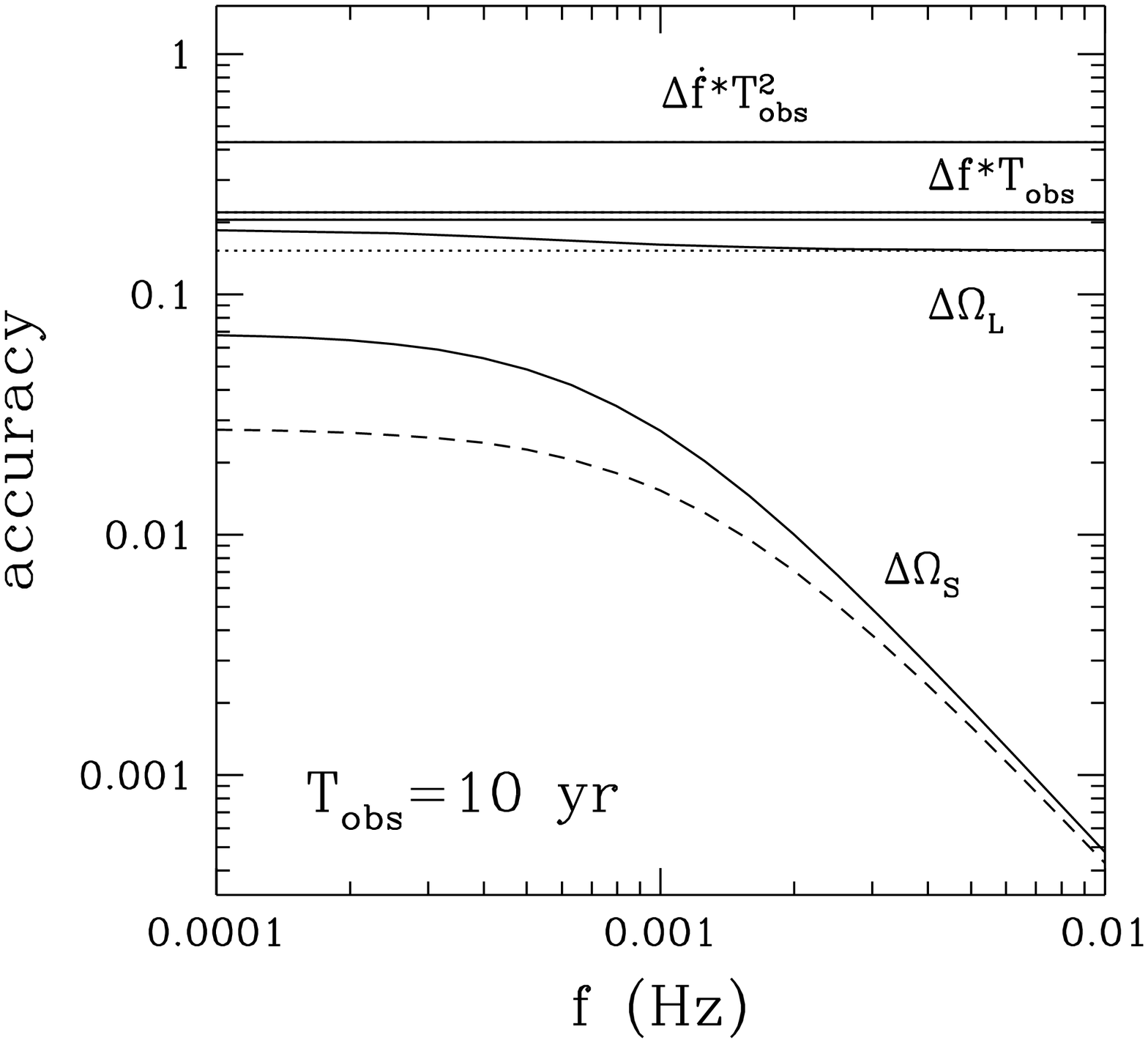}
  \caption{Same as Fig.\ref{f1}, but as a function of the frequency $f$
 with the observational period $T_{obs}=1,10$ yr. The accuracies are
 normalized by $SNR=10$.} 
\label{f2}
\end{figure}

Fig.\ref{f2} is same as Fig.\ref{f1} but given as a function of
frequency  $f$ with fixed observational period
 $T_{obs}=1$ and 10 yr.
We find that for $T_{obs}=10$ yr the errors 
$\Delta A,\Delta f$ and $\Delta \dot{f}$
 do not depend on the frequency,  as given in equations
 (\ref{ga})-(\ref{gc}).
But the angular resolution $\Delta \Omega_S$ depends on the frequency
 and is given  by
\beqa
\Delta \Omega_S &=& 4.8 \times 10^{-4}
 \lmk \frac{f}{10^{-2} \mbox{Hz}} \rmk^{-2}
 \lmk \frac{SNR}{10} \rmk^{-2} \mbox{sr},
\label{fdos}
\eeqa
for higher frequencies $f \gsim 10^{-3}$ Hz, and nearly constant for
 lower frequency $f \lsim 10^{-3}$ Hz.  

Now we discuss correlation between Fisher matrix elements for $T\gsim 2
$yr. As expected
 from Figs 1 and 2,  the source direction  $\Omega_S$ has almost no
correlation with other parameters at higher frequencies $f>10^{-3}$Hz
\footnote{In other words, Fisher information matrix $\Gamma_{ij}$ in 
equation
 (\ref{fis}) becomes diagonalized
 to the two parts; 
$
\Gamma \simeq \left( 
           \begin{array}{@{\,}cc@{\,}}
           \Gamma_A &     0        \\
               0    &  \Gamma_S    \\ 
           \end{array} \right),  
$
where $\Gamma_A$ represents a $6 \times 6$ matrix which corresponds
 to $(A,f,{\dot f},\phi_0,{\bar \theta_L},{\bar \phi_L})$ components 
and
 $\Gamma_S$ represent a $2 \times 2$ matrix which corresponds to
 $(\bar{\theta}_S,\bar{\phi}_S)$ components.}
and weakly correlates with source orientation $\Omega_L$ at lower  
frequencies $f<10^{-3}$Hz.  This dependence  seems reasonable 
considering 
the information used to determine the source direction  $\Omega_S$,
namely, the Doppler phase at $f>10^{-3}$Hz and amplitude modulation at
$f<10^{-3}$Hz. The orientation $\Omega_L$ is  strongly related to
the latter. As a result angular variables $\Omega_S$ and $\Omega_L$
correlate at lower frequencies. 
The detected wave form $h_\alpha$ is modulated by annual revolution
and rotation of the detectors, and its phase has oscillating part due
to the modulation that is more prominent at higher
frequencies. Amplitudes of the derivatives $\p h/\p \Omega_s$ also show
this oscillation.  In  the Fisher matrix their cross terms with
 other derivatives are significantly canceled with long time
 integration. As a result the estimation errors for the angular 
direction  $\Omega_s$  show very weak correlation with other errors.  

In the same manner we can also understand the asymptotic frequency
dependence of the angular resolution $\Omega_S$. In the derivatives $\p
h_{\alpha}/\p \gamma_i$ for the  Fisher matrix
elements in equation (6) the frequency $f$ appears only through the 
Doppler phase
$\phi_D(t)$ in equation (\ref{dop}). At lower frequencies $f<10^{-3}$Hz
 terms from the Doppler
phase ($\propto f^1$) is smaller than the terms from the amplitude
modulation ($\propto f^0$), and consequently  the
estimation error $\Delta \Omega_S$ does not depend on the frequency $f$.
At higher frequencies  $f>10^{-3}$Hz the Doppler phase term  becomes
dominant  and the Fisher matrix are diagonalized. Thus the error
 $\Delta \Omega_S$ depends on the frequency $f$ as 
$\Delta \Omega_S\propto f^{-2}$ (Cutler \& Vecchio 1998, Moore \& 
Hellings 2002).

\begin{figure}
  \hspace{3.75cm}
   \includegraphics[height=7.5cm,clip]{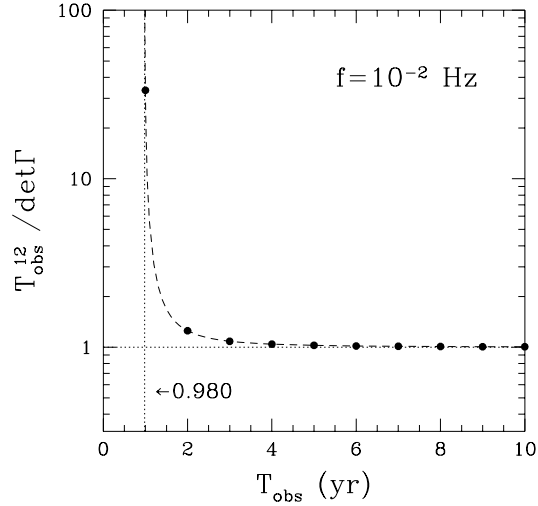} 

\caption{The inverse of the determinant of the Fisher matrix for the
 simple waveform given in equation (16).
The filled circles represent with integer  $T_{obs}$ in units of year.
We normalize the overall scale by unity  at $T_{obs}=+\infty$.
}

\label{f3}
\end{figure}

Finally we analytically investigate how  the estimation errors  improve
 with the observational period $T_{obs}$.  We reanalyze the simple toy 
model
 $h(t)=A \sin[2\pi(f+{\dot f}t/2)t+\phi_0]$ (see the sentences before
 equation (\ref{aa})) adding the information of the source direction
 $(\bar{\theta}_S,\bar{\phi}_S)$ by the Doppler phase  $\phi_D(t)$.
This waveform is given as
\beq
  h(t)=A \sin[~2\pi(f+{\dot f}t/2)t+\phi_D(t)+\phi_0],
\label{swf}
\eeq
with six fitting parameters $\gamma_i=(A,f,{\dot f},\phi_0,
 \bar{\theta}_S,\bar{\phi}_S)$. 
We evaluate the magnitude of the variance-covariance matrix $\langle
 \Delta \gamma_i \Delta 
 \gamma_j \rangle$
 by its   determinant as
  $\det \langle \Delta \gamma_i \Delta \gamma_j \rangle 
 = (\det \Gamma)^{-1}$.
After some algebra the time dependence of $\det \Gamma$ is given
analytically  from
equations 
 (\ref{fis}) and (\ref{swf}) as,      
\beq
 \det \Gamma \propto T_{obs}^{12} \lmk 1- \frac{6}{\pi^2
 \lmk T_{obs}/ 1 \mbox{yr} \rmk^{2}} \rmk 
  \lmk 1- \frac{90}{\pi^4 \lmk T_{obs}/1 \mbox{yr}\rmk^{4}} \rmk,
\label{det}
\eeq       
where we assume that the observational period  $T_{obs}(\gg f^{-1})$ is
 a integer in
units of year.
In  the above expression the quantity  $(\det \Gamma)^{-1}$ formally
  diverges at
 $T_{obs}=\sqrt[4]{90}/\pi=0.98$ [yr] which is very close to 1, thus 
the
 estimation errors have large value at $T_{obs}=1$ yr.  
Fig.\ref{f3} shows  the inverse of the determinant of the Fisher matrix
in the form  $T_{obs}^{12}(\det \Gamma)^{-1}$ as a function of the 
observational
 period $T_{obs}$.
  From Fig.\ref{f3}, $T_{obs}^{12} (\det \Gamma)^{-1}$ rapidly 
converges at
 $T_{obs} \sim 2$ yr.
In this case we also obtain the estimation errors as follows;
\beqa
\frac{\Delta A}{A} &=& \frac{1}{SNR} ,\\ 
\Delta f&=&\frac{4\sqrt3}{\pi}\frac{T^{-1}_{obs}}{SNR} 
 \times \sqrt{\frac{1-(45/8)~x^2 (1+x^2)}{(1-6 x^2)(1-90 x^4)}}, \\
\Delta {\dot f}&=&\frac{6\sqrt5}{\pi}\frac{T^{-2}_{obs}}{SNR}
 \times \frac{1}{\sqrt{1-90 x^4}} , \\
\Delta \Omega_S &=& \frac{1}{\pi f^2 R^2 \cos{\bar{\theta}_S} SNR^2}
 \times \frac{1}{\sqrt{(1-6 x^2)(1-90 x^4)}}  \\
 &\simeq& 4.3 \times 10^{-4} \lmk \frac{\cos{\bar{\theta}_S}}{0.3} \rmk
^{-1}
 \lmk \frac{f}{10^{-2} \mbox{Hz}} \rmk^{-2}
 \lmk \frac{SNR}{10} \rmk^{-2} \mbox{sr}, \label{doms}
\eeqa
where $x=\pi^{-1} (T_{obs}/1 \mbox{yr})^{-1}$ and equation
 (\ref{doms}) is valid for $T_{obs} \gsim 2$ yr and is similar to
 equation (\ref{fdos}).
The above results for $\Delta f$, $\Delta \dot{f}$ and
 $\Delta \Omega_S$ are about $40 \%$ smaller than the the results
 in Fig.\ref{f1} at $T_{obs}=1$ yr, because we do not include the
 information of the source orientation.  
$\Delta f$ and $\Delta \dot{f}$ are asymptotically same as
 equations (\ref{af}) and (\ref{ac}) for $T_{obs} \gsim 2$ yr.

\subsection{Statistical Analysis}
So far we have studied a specific set of the angular parameters $\hat n
$
and $\hat L$. In this subsection we present statistical results for
their various combinations  at the asymptotic region $T_{obs}\gsim2$yr.
 
We have made 100 realizations of  $\hat n$
and $\hat L$ that  are distributed randomly on  celestial
spheres. Then we 
calculate the
estimation errors $\Delta f$ and $\Delta {\dot f}$ for each binary
normalized by $SNR=10$ with $T_{obs}=10$yr.  We find that theses errors
 depend very  weakly (less than 10\% scatter)  on  
the directions  $\hat n$
and $\hat L$ in contrast  to the results for  $T_{obs}=1$yr
 (scattering typically factor $2^{\pm1}$ around the average).
We calculate their mean values and
obtain  results 
given in the following forms
\beqa
\Delta f &=&0.22 \lmk \frac{SNR}{10} \rmk^{-1} T_{obs}^{-1},
\label{nf} \\
\Delta {\dot f} &=&0.43 \lmk \frac{SNR}{10} \rmk^{-1} T_{obs}^{-2}.
\label{nc}  
\eeqa
Note that these   results do not depend on the  frequency $f$ and
would be useful for quantitative analysis of quasi-monochromatic
binaries with $T_{obs}\gsim 2$yr. As expected from  the previous
subsection the simple analytical
 estimations given in  equations (\ref{af})
 and (\ref{ac})
are  very close to equations (\ref{nf}) and (\ref{nc}).

We have also studied the estimation  errors $\Delta \Omega_S$, 
$\Delta \Omega_L$ and $\Delta A/A$.   Distributions of  the latter two
have very large scatters.  This is because (i) the Fisher matrix 
elements
relating to the orbital orientation $\hat L$ become singular at the  
highly symmetric face-on configuration ${\hat n}=\pm {\hat L}$ and (ii)
 the
estimation of the amplitude $A$
is closely related to the inclination.  For majority of realizations
with  $|{\hat n}\cdot {\hat L}| \lsim 0.8$  we have typically
 $\Delta A/A\sim 0.2 (SNR/10)^{-1}$. To determine the source
direction $\hat n$ we can use the information of the Doppler phase in
addition to the annual amplitude modulation caused by LISA's rotation. 
Thus the error $\Delta \Omega_S$ does not show such a bad behavior. For
the above 100 realization at $f=0.01$Hz  with $SNR=10$ and $T_{obs}=10$
yr
 the errors $\Delta
\Omega_S$ are distributed as  $1.3\times 10^{-4}{\rm sr}\le \Delta
\Omega_S \le 3.7 \times 10^{-3}$sr with the mean value $\Delta
\Omega_S =7.1 \times 10^{-4}$sr. Therefore the following relation 
roughly
gives  the estimation error for the
 source direction with $T_{obs}\ge 2$yr  
\beq
\Delta
\Omega_S \sim 7.1 \times 10^{-4} \lmk \frac{SNR}{10} \rmk^{-2}\lmk
\frac{f}{10^{-2}\rm Hz} \rmk^{-2} {\rm sr},
\eeq
at $f\gsim 2\times 10^{-3}$Hz where the Doppler phase becomes more 
important
than the annual amplitude due to the rotation of LISA.

Now let us discuss  issues concerned with the chirp signal $\dot f$
based on 
 equation (\ref{nc}). We  assume that the chirp signal is dominated by
gravitational radiation reaction  and introduce a  parameter
 $R\equiv \Delta {\dot
f}/ {\dot f}$ that represents relative accuracy  of the
measured signal $\dot f$.  For a given threshold $R$ we can solve the
corresponding frequency $f_R$ as
\beq
f_R=9.2\times 10^{-4} \lmk \frac{M_c}{1M_\odot}\rmk^{-5/11}
\lmk\frac{SNR}{10} \rmk^{-3/11}  \lmk \frac{T_{obs}}{10 {\rm yr}}
\rmk^{-6/11} 
\lmk \frac{R}{1.0}\rmk^{-3/11}  ~~{\rm Hz} 
\eeq
for observational period  $T_{obs}\gsim 2$yr.
The frequency  $f_{R=1}$ can be regarded as the critical frequency for 
treatment of the
chirp signal $\dot f$. At higher frequencies $f\gsim f_{R=1}$ the
parameter 
$\dot f$ should be included in the matched filtering,  but the simple
prescription 
$\dot f=0$  
would be better  at  $f\lsim f_{R=1}$ since the expected signal would 
be
completely buried  in error.  If  the chirp signal is measured  with
accuracy $R(\ll 1)$,  the chirp mass can be estimated with relative
accuracy $\Delta M_c/M_c\simeq 3R/5$ from relation ${\dot f}\propto
M_c^{5/3}$. 
The chirp signal $\dot f$  is also  essential to
determine the distance $D$ to the binary. From the
simple relation $A=5{\dot f}/96\pi^2f^3D$,  
the estimation error for distance $D$ is roughly evaluated as
\beqa
  \frac{\Delta D}{D} &\simeq& \frac{\Delta A}{A}+\frac{\Delta {\dot f}}
 {\dot f},  \\
&\simeq& {\rm Max}\lnk \frac{\Delta A}{A}, ~~~\frac{\Delta {\dot f}}
 {\dot f}\rnk,  \\
  &\simeq& 0.2 \left( \frac{SNR} {10} \right)^{-1} {\rm Max} \lnk 1,
~~~\lmk\frac{f}{1.4\times 10^{-3}{\rm Hz}} \rmk^{-11/3}
\lmk\frac{M_c}{1 M_\odot} \rmk^{-5/3}
\lmk\frac{T_{obs}}{10{\rm yr}} \rmk^{-2}   \rnk,  
\label{dis}
\eeqa
where we have used the typical error for the amplitude estimation
$\Delta A/A\sim 0.2(SNR/10)^{-1}$. For  compact binaries with chirp 
mass $M_c\sim
1M_\odot$ such as NBs or CWDBs  the chirp signal $\dot f$ is the 
dominant
source of the error    at  lower frequencies $f\lsim
10^{-3}$Hz and observational period $T_{obs}\sim 10$yr.   The amplitude
$A$ becomes dominant one  at  $f\gsim 
10^{-3}$Hz.

Finally, we calculate the estimation error for the three dimensional
 position of the
 Galactic binaries. 
The signal to noise ratio is calculated from equation (\ref{snr})
 as $SNR \sim 380 
 ~( f/5 \times 10^{-3} \mbox{Hz} )^{2/3}$
 $( D/10 \mbox{kpc} )^{-1} 
 ~( T_{obs}/10 \mbox{yr} )^{1/2}
 ~( M_c/1 M_{\odot} )^{5/3}
 ~( \sqrt{S_n}/ 8 \times 10^{-21} \mbox{Hz}^{-1/2} )^{-1/2}$.
 The noise spectrum $S_n(f)$ (in units of ${\rm Hz}^{-1}$) is nearly
 constant for $3 \times 10^{-3}~\mbox{Hz} \lsim f \lsim 10^{-2}$ Hz
 (see Fig.5 in Cutler 1998).
Hence one could determine both the angular position and the distance
 to the binary with the accuracy of
\beqa
 \Delta {\bar \theta_S} &\sim& 2~
 \left( \frac{f}{5 \times 10^{-3} \mbox{Hz}} \right)^{-5/3}
 \left( \frac{D}{10 \mbox{kpc}} \right)
 \left( \frac{T_{obs}}{10 \mbox{yr}} \right)^{-1/2} \nonumber \\
 && ~~~~~~~\times \left( \frac{M_c}{1 M_{\odot}} \right)^{-5/3}
 \left( \frac{\sqrt{S_n}}{8 \times 10^{-21} \mbox{Hz}^{-1/2}}
 \right)^{1/2} \mbox{arcmin},\\
 \frac{\Delta D}{D} &\sim& 5 \times 10^{-3}
 \left( \frac{f}{5 \times 10^{-3} \mbox{Hz}} \right)^{-2/3}
 \left( \frac{D}{10 \mbox{kpc}} \right)
 \left( \frac{T_{obs}}{10 \mbox{yr}} \right)^{-1/2} \nonumber \\
 && ~~~~~~~\times \left( \frac{M_c}{1 M_{\odot}} \right)^{-5/3}
 \left( \frac{\sqrt{S_n}}{8 \times 10^{-21} \mbox{Hz}^{-1/2}} \right)^{
1/2}.
\eeqa

\section{Conclusion}
We have calculated LISA's measurement accuracy for  short-period 
binaries
 $(10^{-4}~{\mbox Hz} \lsim$ $f$ $\lsim 10^{-1}~{\mbox Hz})$
 in our  Galaxy, including the effects of chirp signal ${\dot f}$
 and dependence  on observational period $T_{obs}$.
We find that the measurement accuracy  rapidly improves
 for $T_{obs} \gsim 2$yr comparing with $T_{obs}\sim 1$yr. This might 
be
an important element  for discussing  operation period of LISA.

At observational period 
$T_{obs} \gsim 2$ the errors for quantities  $f$ and ${\dot f}$
 is independent on the information of the positions and orientations of
binaries  in contrast to 
 $\Delta A$ and $\Delta \Omega_{S,L}$.
It is also found that the estimation errors $\Delta A$, $\Delta f$ and
$\Delta{\dot f}$ 
 are almost independent on the frequency $f$. 
The fitting formulae of the estimation errors for the source
 parameters (such as frequency $f$, chirp signal ${\dot f}$,
 amplitude $A$, angular position $\Omega_S$ and distance $D$)
 are given as functions of $f$ and $T_{obs}$. We expect
these would be  powerful tools for quantitatively studying possibility
 gravitational wave astronomy.

The relative motion between the Sun and a binary could affect the
 frequency $f$ by the Doppler factor, $(1+v_R)$ where 
 $v_R$ is the relative radial velocity.
The tangential velocity $v_T$ or accelerating motion $\dot v$  would
affect  
 the time variation of the frequency, $(\dot{f}/f)_{gal}  \sim
v_{T}^2/cD$ or $\sim {\dot v}/c\sim
v_{rot}^2/cR_g$ respectively 
 where $v_{rot}$ is the rotation velocity ($\sim 200$km/s) of the
Galaxy, $R_g$ is the  Galactic radius $\sim 10$kpc and $D$  
 is the distance to the binary (see {\it e.g.} Damour \& Taylor 1991). 
For most Galactic binaries in LISA band  this should be negligibly 
small
compared with  
 $(\dot{f}/f)_{GW}$ due to gravitational radiation reaction.
This condition is expressed as $f \gg 3\times10^{-5}
 (M_c/1M_{\odot})^{-5/8} (v_{rot}/200 \mbox{km})^{3/4}
 (D/10\mbox{kpc})^{-3/8}$ Hz.
For  binaries very close to us  $D \lsim 100$ pc   
 these effects would be important.  But note that we might measure the
proper 
motion of such binaries by optical observation.

\acknowledgments
We would like to thank an anonymous referee for helpful comments to
improve the manuscript,  and Takeshi Chiba, Hideyuki Tagoshi and 
Hirotaka
Takahashi  for useful comments and discussions.
This work was
supported in part by Grant-in-Aid of Scientific Research
of the Ministry of Education, Culture, Sports, Science and
Technology No. 0001416.

\newpage

\begin{table}
  \begin{center}
  \setlength{\tabcolsep}{10pt}
  \renewcommand{\arraystretch}{1.1}
  \begin{tabular}{cccccc} \hline\hline
   $T_{obs}$ (yr) & ~~$\Delta A/A$~~ & ~$T_{obs}\Delta f$~ & ~$T_{obs}^
2\Delta {\dot f}$~ & ~$\Delta \Omega_S$ (sr)~ & ~$\Delta \Omega_L$ (sr)
 \\ \hline
  \multicolumn{6}{l}{$f=10^{-4}$ Hz} \\ 
   1yr & 0.205 & 0.33 & 0.59 & 8.47 $\times 10^{-2}$ &
   0.201 \\ 
& 0.204 & 0.076 &   & 8.27 $\times 10^{-2}$ &
   0.199 \\ 
       & 0.204 &  0.31 & 0.58 & & 0.154 \\
   10yr & 0.204 & 0.22 & 0.43 & 6.78 $\times 10^{-2}$ &
   0.185 \\ 
& 0.204 & 0.055 &   & 6.78 $\times 10^{-2}$ &
   0.185 \\ 
        & 0.204 & 0.22 & 0.43 & & 0.152 \\  \hline
  \multicolumn{6}{l}{$f=10^{-3}$ Hz} \\ 
   1yr & 0.205 & 0.62 & 1.1 & 7.59 $\times 10^{-2}$ &
   0.185 \\
& 0.204 & 0.092 &  & 3.97 $\times 10^{-2}$ &
   0.169 \\     
        & 0.204 &0.31 & 0.58 & & 0.154 \\
   10yr & 0.204 & 0.22 & 0.43 & 2.70 $\times 10^{-2}$ &
   0.161 \\
        & 0.204 &0.055 &  & 2.70 $\times 10^{-2}$ & 0.161\\
 & 0.204 & 0.22 & 0.43 &  &
   0.161 \\ \hline
  \multicolumn{6}{l}{$f=10^{-2}$ Hz} \\
   1yr & 0.205 &1.1 &2.2 &  
    4.10 $\times 10^{-3}$ & 0.155 \\
& 0.204 &0.14 &0.20 &  
    1.08 $\times 10^{-3}$ & 0.153 \\
       & 0.204 & 0.31 &0.58 & & 0.153 \\
   10yr & 0.204 &0.22 &0.43 &
   4.77 $\times 10^{-4}$ & 0.110 \\
 & 0.204 &0.055 & &
   4.77 $\times 10^{-4}$ & 0.153 \\
       & 0.204 &0.22 &0.43 & &
   0.109 \\ \hline\hline
  \end{tabular}
  \end{center}
\caption{
LISA's measurement accuracy for  binaries with angular parameters
 $(\cos \bar{\theta}_S=0.3, \bar{\phi}_S=5.0, \cos \bar{\theta}_L=-0.2,
 \bar{\phi}_L=4.0)$.   
Results are normalized by $SNR=10$. Errors scale as $(SNR/10)^{-1}$ for 
 $\Delta A,\Delta f$ and $\Delta {\dot f}$, and $(SNR/10)^{-2}$ for
 $\Delta \Omega_{S,L}$.
The second lines in each observational period  $T_{obs}$ represent the 
case with removing the
 chirp signal $\dot f$ from fitting parameters and the third lines with
 removing the direction of the source $(\theta_S, \phi_S)$.  }
\label{t1}
\end{table}

\end{document}